\documentclass[aps,prl,comment,twocolumn,superscriptaddress,showpacs,floatfix]{revtex4}
\usepackage{epsfig}
\newcommand{\be}{\begin{equation}}
\newcommand{\ee}{\end{equation}}
\newcommand{\bc}{\begin{center}}
\newcommand{\ec}{\end{center}}
\newcommand{\bi}{\begin{itemize}}
\newcommand{\ei}{\end{itemize}}
\newcommand{\ba}{\begin{eqnarray}}
\newcommand{\ea}{\end{eqnarray}}

\newcommand{\ignore}[1]{}
\begin{document}
{\bf Comment on ``Critical branching captures activity in living
neural networks and maximizes the number of metastable states''}.\\

In a recent Letter, Haldemann and Beggs \cite{HB2005} use a
branching process to simulate propagated neuronal activity in form
of neuronal avalanches.  This work built on an experimental paper
by Beggs and Plenz \cite{BP2003}, which demonstrated that a
critical branching process captures some of the dynamics of
propagation of neuronal activity through synchronized groups of
neurons. Experimentally the branching parameter $\sigma$ is
measured as the ratio of the number of ``descendant'' electrodes
to the number of ``ancestor'' electrodes activated in each
avalanche. It was found in \cite{BP2003} that, under normal
activity, cortical networks exhibit scale-free avalanches and
$\sigma= 1$. In the recent Letter \cite{HB2005}, the authors
stated that the experiments reported in \cite{BP2003} exhibited
$\sigma > 1$ during epileptic activity. This statement indicating
that epileptic networks have  $\sigma  > 1$ is essential to the
authors hypothesis and conclusions, but is false. First, the
referenced paper \cite{BP2003} does not provide any information
about the experimentally obtained branching parameter in epileptic
networks and second, (as shown in Fig. 1) the branching parameter
in epileptic networks is {\em smaller} and not larger than 1.
\begin{figure}[ht]
\centering \psfig{figure=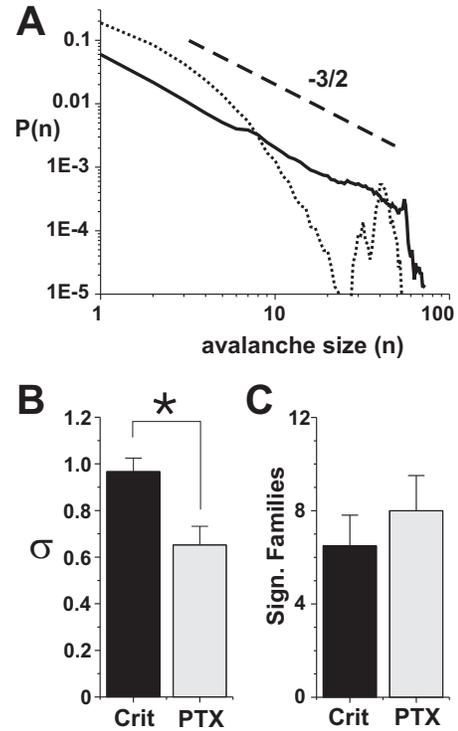,width=2.3
truein,clip=true,angle=0} \caption{\footnotesize{Experimental
estimate of branching ratio and number of different patterns in
networks for two conditions: critical and made epileptic by PTX.
{\bf A.} Representative examples of avalanche size distribution in
critical and epileptic networks ( solid and dotted line
respectively). {\bf B.}
  The branching ratio $\sigma$ $\sim 1$ in the
critical state \cite{BP2003} (``Crit'') and smaller than one in
epileptic networks (``PTX'') \cite{note1}. {\bf C.}The number of
metastable states is not different for critical and epileptic
networks \cite{note2}. }}
\end{figure}
\\
Fig. 1 summarizes information on cultured cortical networks
including those networks used in \cite{BP2003}, that had been made
epileptic by blocking fast inhibition using the antagonist
picrotoxin (PTX). As reported previously \cite{BP2003}, networks
in the critical state reveal a scale free avalanche size
distribution, but when neuronal inhibition is blocked with drugs
such as PTX a characteristic avalanche size appears and the
distribution becames bimodal (Fig. 1A). Under these conditions,
the branching parameter $\sigma$, calculated according to the
formula found in \cite{BP2003}, is significantly smaller than that
for the critical state (Fig. 1B), which is in contrast to what has
been stated by Haldemann and Beggs \cite{HB2005}.  This result
holds true for all three described approaches of calculating
$\sigma$ in these networks. Similarly, the number of significant
families, i.e. the metastable states discussed in \cite{HB2005},
is not different for the critical and epileptic state in the real
cortical networks (Fig. 1C), which is, again, in contrast to the
claims by Haldemann and Beggs \cite{HB2005}.\\

In conclusion,  the first claim in the Letter of Haldeman and
Begss stating that ``...the model mimicked the double peaked
distribution produced when we bathed the cortical cultures in
picrotoxin, an agent that selectively blocked inhibition and
increased $\sigma$.'' is misleading. The analysis of that data
shows otherwise: as seen in Fig.  1B $\sigma$ {\em decreases}. The
second claim in Haldeman and Begss indicating that cortical
cultures showed increased number of metastable states is also
false: as shown in Fig 1C there is no significant difference
between the number of states in critical and epileptic networks.

Dietmar Plenz. \\
{\footnotesize{Unit of Neural Network Physiology,
Laboratory of Systems Neuroscience, National Institute of Mental
Health, 35 Convent Drive,  Bethesda, Maryland, USA.}}


\begin{thebibliography}{999}
\bibitem{HB2005}C. Haldeman and J. M. Beggs. Phys. Rev. Lett. {\bf94}, 058101 (2005).
\bibitem{BP2003}J. M. Beggs and D. Plenz. J. Neurosci. {\bf23}, 11167-11177 (2003).
\bibitem{BP2004}J. M. Beggs and D. Plenz. J. Neurosci. {\bf24}, 5216-5229.(2004).
\bibitem{note1} $\sigma$ = 0.97 $\pm$ 0.15 for critical, and 0.65 $\pm$ 0.18
for PTX networks (mean $\pm$ SEM); dF1,10 = 10.75, P = 0.008;
ANOVA. Calculated from 5 PTX  and 7 critical networks ($\sim$
40,000 avalanches and 10 hrs recording per network and state).
\bibitem{note2}  Sign. Families = 6.49 $\pm$ 1.3 for critical, and 8.0 $\pm$ 1.5
for PTX networks (mean $\pm$ SEM); dF1,10 = 0.56; ANOVA; P = 0.47.
Calculated from 5 PTX  and 7 critical networks (1000 consecutives
avalanches from each), repeated 5 times and averaged. Significance
was established with 20 rate-matched shuffle sets at Type I error
= 5 \% as in \cite{BP2004}.
\end{thebibliography}
\end{document}